# Quantum Trapping on MoS$_2$ through Lifetimes of Resonant Electrons: Revealing Pauli Exclusion Principle

Shin-Ming Lu,[1] Wei-Bin Su,[1*] Horng-Tay Jeng,[2,1,3*] Wen-Yuan Chan,[1] Ho-Hsiang Chang,[1] Woei Wu Pai,[4] Hsiang-Lin Liu,[5] and Chia-Seng Chang[1]

[1]*Institute of Physics, Academia Sinica, Nankang, Taipei 11529, Taiwan*

[2]*Department of Physics, National Tsing Hua University, Hsinchu 30013, Taiwan*

[3]*Physics Division, National Center for Theoretical Sciences, Hsinchu 30013, Taiwan*

[4]*Center for Condensed Matter Sciences, National Taiwan University, Taipei 10617, Taiwan*

[5]*Department of Physics, National Taiwan Normal University, Taipei 11677, Taiwan*

We demonstrate that the linewidth of the field emission resonance (FER) observed on the surface of MoS$_2$ using scanning tunneling microscopy can vary by up to one order of magnitude with an increase in the electric field. This unusual phenomenon originates from quantum trapping, where the electron relaxed from the resonant electron in the FER is momentarily trapped in the potential well on the MoS$_2$ surface due to the wave nature. Because the relaxed electron and the resonant electron have the same spin and the action of the Pauli exclusion principle, the lifetimes of the resonant electrons can be substantially prolonged when the relaxed electrons engage in the resonance trapping. The linewidth of the FER is thus shrunk considerably to as narrow as 12 meV. Moreover, it was found that the valley intensities around the FER are zero, indicating that MoS$_2$ has an energy gap above the vacuum level. By using the energies of FERs, the potential of the band bending in the MoS$_2$ interior can be measured precisely.

In quantum mechanics, two cases are considered in the model of a square potential well. They are quantum confinement and quantum scattering. In reality, these two cases can be observed in metallic films whose electronic structures have the free-electron property [1−10]. Herein, we report a third case called quantum trapping, where the particle is in the well but has energy higher than the well depth. In this case, the particle can be trapped in the well temporarily; however, it can eventually escape from the well because of wave function dissipation. The mean duration for which the well traps the particle varies with the well size in oscillation. We demonstrate that quantum trapping can be experimentally revealed through the linewidth of the field emission resonance (FER) [11,12] observed on the surface of bulk $MoS_2$ using scanning tunneling microscopy (STM). We discovered that the linewidth oscillates with the electric field of generating FER. Because the inverse of the linewidth can reflect the mean lifetime of resonant electrons that remain in the FER, we suggest that the oscillatory trapping duration is manifested by this variation of the lifetime. The relationship between them is established by following mechanisms. (1) The lifetime of a resonant electron terminates at the occurrence of electron relaxation via light emission [13,14]. (2) Because of the inhomogeneous electric field resulting from the curvature of the STM tip, the relaxed electron has lateral kinetic energy [15] and a local potential well forms beneath the tip on the $MoS_2$ surface. Due to quantum trapping, the potential well temporarily traps the relaxed electron. (3) Both the relaxed electron and the resonant electron have the same spin. According to Pauli exclusion principle, the resonant electron cannot emit light while the relaxed electron remains trapped in the well.

Although FERs originate from the quantized states in a vacuum [16], they can contain the information associated with physical properties of the surface and the STM tip. For example, FER energies can reveal the local work function of the surface [17−23]. FER intensities can reflect the local electron transmissivity of the reconstructed surface [24]. The zero valley intensities appearing around the FER may indicate that the observed material has a band gap above the vacuum level [24]. The number of FERs can manifest the sharpness [25,26] and field enhancement factor [27] of an STM tip. Previous studies have demonstrated that the FER can be used to investigate the atomic structure of the insulator [28], plasmon-assisted electron tunneling [29], and the dynamics [15,30] and lateral quantization [31,32] of surface electrons above the vacuum level. In this study, we focus on the linewidth of the FER, which has seldom been investigated. The inverse of the linewidth can be interpreted as the lifetime of the electron staying in the FER state referring to a similar interpretation for quantum well states in metallic films [4].

Figure 1(a) shows a differential $Z$-$V$ spectrum with five FER peaks observed on a $MoS_2$ surface whose topographic image is displayed in the inset. The numbers represent the orders of FER peaks. A previous study noted that the potential of the zeroth-order peak is the superposition of the image potential and the external potential, whereas that of high-order peaks is simply the external potential [17]. Thus, the high-order numbers are also the quantum numbers of quantized states in the external potential. The dashed line in Fig. 1(a) indicates the zero spectral intensity. The intensities of valleys (marked by arrows) around the peak of FER 2 are exactly zero, implying that $MoS_2$ has a band gap above the vacuum level and the energy of FER 2 is in this band gap [24].

Figure 1(b) displays the FERs with the numbers of three, four, and six, respectively. Our recent study explained that a larger number of FERs results from a sharper STM tip [25,26]. Figure 1(b) also shows that the high-order peaks in the spectrum with three FERs are all much narrower than those in the spectra with four and six FERs. This phenomenon inspired us to explore the physics behind the linewidth of the high-order FERs. Because of zero valley intensities, the peak of FER 2 is the most appropriate for analyzing the linewidth. Figures 1(c)-1(f) depict the Lorentzian fittings of FER 2 in Figs. 1(a) and 1(b). The linewidths extracted from the fittings change with the FER numbers in oscillation, and the variation can be up to one order of magnitude. In particular, the linewidth can be as narrow as 12 meV for the case of five FERs.

If an FER is a result of the quantized state in the triangular potential, its energy $E_n$ obeys:

$$E_n = E_{vac} + \alpha F^{\frac{2}{3}}(n - \frac{1}{4})^{\frac{2}{3}}, \qquad (1)$$

where $E_{vac}$ is the vacuum level, $F$ is the electric field, n=1, 2, 3…, is the quantum number, and $\alpha=(\hbar^2/2m)^{1/3}(3\pi e/2)^{2/3}$. Figure 2(a) shows plots of the energies of the high-order peaks in Figs. 1(a) and 1(b) versus $(n-1/4)^{2/3}$. The results show that the data points in the cases of four, five, and six FERs can be fit well by the lines, indicating that equation (1) is valid. From the line slope and the extrapolated value, one can obtain $F$ and $E_{vac}$, respectively. Figure 2a reveals that a higher number of FERs corresponds to a weaker $F$. Therefore, the oscillatory linewidth is related to the electric field for generating FERs.

Figure 2(b) illustrates that an STM tip mainly consists of a base with a radius (typically a few tens of nanometers) and a protrusion (marked by an arrow) [26]. The sharpness of an STM tip is determined by the open angle of the protrusion. $F$ obtained from Fig. 2(a) appears at the dashed line normal to the surface for a range from the surface (o point) to the classical turning point (cross) of the resonant electron. Because MoS$_2$ is a dielectric material with a dielectric constant of 3.7 [33], the electric field can penetrate its surface to cause band bending in the interior [34]. Thus, the electric potential along the surface normal is $V_b+Fz$, where $V_b$ is the potential of band bending and $z$ is the distance from surface. The electric potential below the classical turning point in the lateral direction can expressed as [15]

$$U_L(\rho, z) = \frac{(V_b+Fz)}{[1+\left(\frac{\rho}{\rho_0}\right)^2]}, \qquad (2)$$

where $\rho_0$ is a constant and $\rho$ is the lateral distance from the dashed line, revealing the distribution curve in Fig. 2(b).

Figure 2(c) displays the projected bulk band structure of MoS$_2$ obtained by DFT calculations [35–39] for energy within the range of 1–12 eV above the Fermi level. A band gap appears above the vacuum level (5.2 eV) [40] that has the band edges at 6.1 and 8.2 eV at the gamma point, marked by lines 2 and 3. Thus, the calculation result is consistent with the observation of zero valley intensity. However, the energy of FER 2 in Fig. 1(a) is 8.5 eV (dashed line) outside of the band gap. This inconsistency can be attributed to band bending because the vacuum level determined from the extrapolation in Fig 2(a) is 6.45 eV for the case of five FERs. The potential energy of band bending is determined to be 1.25 eV (6.45 eV-5.2 eV) accordingly. The energy diagram on the right-hand side of Fig. 2(c) illustrates that the band edges and the vacuum level (line 1) in the bulk all move upward when approaching the surface due to band bending. Therefore, the band edges at the surface are 7.35 and 9.45 eV. Consequently, the energy of FER 2 is in the band gap.

In the FER, resonant electrons move back and forth within a specific distance between the surface and the classical tuning point. Because $E_n$, $F$, and $E_{vac}$ are known, this distance $s$ corresponding to an FER can be calculated as $(E_n-E_{vac})/eF$. Subsequently, the round-trip (RT) time $t$ for resonant electron motions can be calculated from $2s=eFt^2/m$. Figure 2(d) shows the RT time versus $F$ in the case of FER 2, where the RT time is shown on the femtosecond scale. Previous studies have demonstrated that resonant electrons can be relaxed through light emission [13,14] to release quantized energies; however, before relaxation, electrons should at least complete a round trip to form the standing wave that is necessary to reveal the FER. Therefore, the RT time can basically reflect the lifetime of the resonant electron. Thus, Fig. 2(d) shows that the lifetime should monotonically decrease with an increased $F$. However, this finding is evidently contradictory to the observation in Fig. 1 that the linewidth oscillates with the electric field. The oscillation range can be up to one order of magnitude, implying that under a certain electric field, the lifetime can be prolonged substantially. Because the lifetime of the resonant electron terminates at light emission, this significant extension of the lifetime indicates that resonant electrons cannot emit light to remain in the FER for many more round trips. The prohibition of light emission results from the mechanisms (2) and (3). Next, we explain how these mechanisms are valid.

The penetration of the electric field can induce the positive polarization charge on the $MoS_2$ surface. By using equation (2), the density distribution of the polarization charge can be calculated (see Supplemental Material [41]). An electron on the surface under an STM tip is repelled by the electric field in vacuum but is also simultaneously attracted by the polarization charge. Consequently, the electric potential on the surface is the superposition of $U_L(\rho,0)$ and the potential $U_p(\rho)$ due to the polarization charge. Because $U_L$ and $U_p$ are both zero when $\rho$ is infinite, whether the electron faces a potential well can be judged from the sign of the resultant potential $U_r$ at the o point, that is, $U_L(0,0)+U_p(0)$ (see Supplemental Material [41]). Figure 2(e) shows $U_r$ versus $\rho_0$ under electric fields obtained from Fig. 2(a). The results indicate that $U_r$ is negative for $\rho_0$ higher than 18 Å. $\rho_0$ is generally related to the tip base with a radius of tens of nanometers; therefore, it is plausible that $\rho_0$ is higher than 18 Å. Accordingly, $U_r$ is negative, and a potential well exists.

Because the resonant electrons are hot electrons, the lateral kinetic energy $E$ of relaxed electrons is higher than the vacuum level of $MoS_2$, indicating that relaxed electrons have energy higher than the well depth. Therefore, the situation differs from the case of quantum confinement. One can expect that relaxed electrons are trapped to move back and forth in the well because of the formation of standing waves in the radial direction. However, these standing waves would be dissipated because the wave function outside the well is a plane wave instead of an evanescent wave. Thus, relaxed electrons leave the well eventually. To model this case of quantum trapping, we assume that the potential well is a cylindrical square well in which the radial wave function is a Bessel function of the first kind. We note that Bessel functions have oscillatory features similar to cosinusoidal and sinusoidal functions; this enables us to discuss quantum trapping by using a one-dimensional square well with depth $V_0$ and width $2L$ [Fig. 3(a)]. The wave vector $k$ of the relaxed electron in the well satisfies $kL=\beta L(1+E/V_0)^{1/2}$, where $\beta=(2mV_0)^{1/2}/\hbar$. The formation of the standing waves requires the relaxed electrons be reflected at $-L$ and $L$ to move through a complete cycle in the well. If the energy of relaxed electrons is much higher than the depth ($E>>V_0$), they behave similar to classical particles and leave the well without reflection. The standing waves should vanish. Accordingly, the wave functions of relaxed electrons after the first cycle can be $\psi_1(x)=(L)^{1/2}\exp(-\gamma E)\cos(kx)$ or $\psi_2(x)=(L)^{1/2}\exp(-\gamma E)\sin(kx)$, where $\gamma$ is a constant. The exponential term can satisfy no standing wave for $E >> V_0$ because $E$ can be considered infinite under this condition. Thus, the probability in the well is

$$P_1 = \int_{-L}^{L} \psi_1^2 dx = e^{-2\gamma E}\left(1 + \frac{\sin 2kL}{2kL}\right) \quad (3)$$

or

$$P_2 = \int_{-L}^{L} \psi_2^2 dx = e^{-2\gamma E}\left(1 - \frac{\sin 2kL}{2kL}\right). \quad (4)$$

When $E=0$ and $0<2kL<\pi$, $\sin 2kL$ is positive, and thus $P_1>1$ and $P_2<1$. Because the probability should not be higher than 1, the wave function is $\psi_2(x)$ in this case. Similarly, the wave function is $\psi_1(x)$ when $E=0$ and $\pi<2kL<2\pi$.

The results in Fig. 2(e) indicate that $V_0$ and $\beta$ can increase with $F$. Moreover, the area with the polarization charge is larger under a stronger electric field. Hence, $L$ is wider when $F$ is higher. Consequently, the well size $2LV_0$ can be tuned by $F$. Based on equation (2), the electrostatic force in the radial direction enhances with $F$, causing $E$ to also increase with *F*. Therefore, it is plausible to calculate the probability under the conditions that $E/V_0$ is a constant independent of *F* and *E* is replaced with $\beta L$. By combining equations (3) and (4), the probability *P*(1) that the relaxed electron exists in the well after the first cycle can be expressed as

$$P(1) = e^{-2\gamma\beta L}\left(1 - |\frac{\sin 2kL}{2kL}|\right). \quad (5)$$

Therefore, the probability of leaving the well is 1-*P*(1), which can be viewed as the decay rate *R* of the probability. Accordingly, the probability that a relaxed electron stays in the well for *m* cycles *P*(*m*) is $(1-R)^m$. Subsequently, the probability that the relaxed electron has remained in the well for (*m*-1) cycles but leaves the well at the $m^{th}$ cycle is *P*(m-1)-*P*(m). Thus, the average number of cycles $\bar{m}$ of all relaxed electrons is

$$\bar{m} = \sum_{m=0}^{m=m_{max}} m[P(m) - P(m+1)], \quad (6)$$

where $m_{max}$ is the maximum number of cycles and depends on the number *N* of all relaxed electrons when observing the FER.

Equation (5) indicates that *P*(1) is dependent only on $\beta L$; this enables us to show the decay rate versus $\beta L$. Figure 3(b) demonstrate that the decay rate varies with $\beta L$ in oscillation and the local minimum appears when 2*kL* equals integral multiples of $\pi$. The inset in Fig. 3(b) shows $m_{max}$ as a function of the decay rate, revealing that $m_{max}$ increases with a decrease in the decay rate. We use the results in Fig. 3(b) and equation (6) to calculate the average number of cycles $\bar{m}$ versus $\beta L$. Figure 3(c) shows that $\bar{m}$ oscillates with $\beta L$ and the maxima occur exactly at the minima of the decay rate. In addition, $\bar{m}$ is insensitive to *N*.

The time per cycle, equal to $4L/(\hbar k/m)$, is proportional to $L/\beta$. Because both $\beta$ and *L* increase with *F*, it is plausible that the cycle time is independent of *F*. Because $\beta L$ is proportional to the well size, Fig. 3(c) can be interpreted as an indication that the mean duration for which the square well traps the relaxed electrons oscillates with the well size. For certain sizes, the mean trapping time (MTT) is much longer, manifesting resonance trapping. MTT can change by up to one order of magnitude at the first and second resonances marked by 1 and 2 in Fig. 3(c). Evidently, the variation of the linewidths in Figs. 1(c)−1(f) has the characteristics of quantum trapping, leading us to suggest that the lifetime of the resonant electron is governed by the relaxed electron through the Pauli exclusion principle. Thus, MTT is recorded in the linewidth of the FER. The linewidths of 12 meV and 16 meV in Figs. 1(d) and 1(f) result from the relaxed electrons engaging in resonance trapping.

The linewidth of 12 meV corresponds to a lifetime of 27 fs if $\Delta E \Delta t = \hbar/2$. Therefore, the intervention of the Pauli exclusion principle requires the resonant electron and the relaxed electron to coexist on the femtosecond time scale. However, the set current in our experiment is 10 pA equal to one electron per 16 ns. If electrons are emitted one by one every 16 ns, two electrons cannot coexist on the femtosecond scale. In this context, we suggest that every 32 ns, two resonant electrons are emitted into the quantized state in the FER and should have opposite spins because of Pauli exclusion principle [Fig. 3(d)]. One resonant electron is relaxed first through light emission and its spin should be flipped to maximize the total spin according to Hund's rule. Consequently, the resonant electron and relaxed electron can simultaneously exhibit the same spin [Fig. 4(e)], which is similar to the triplet excited state in the phosphorescence.

In summary, we discover that the linewidth of FER is a meaningful quantity that is not only associated with the lifetime of the resonant electron but also able to store information about the behavior of the relaxed electrons. The delivery of information is through the action of the Pauli exclusion principle. In this study, when the size of the potential well situates the relaxed electrons in resonance trapping, the resonant electrons can receive this information to considerably extend their lifetimes.

Therefore, ultra narrow FERs are manifested. Furthermore, the potential well induced by the STM tip should generally be formed on the surfaces of different materials, implying that the phenomenon of quantum trapping can be observed universally. The variation in the well size with the electric field may depend on the dielectric properties of the materials. Therefore, quantum trapping manifested in STM configuration can be developed as a technique for probing the dielectric properties on the nanometer scale.

The authors acknowledge T. K. Lee for helpful discussions. This work was supported by the Ministry of Science and Technology (grant numbers: MOST 106-2112-M-001-004-MY3 and MOST 106-2112-M-007-012-MY3) and Academia Sinica (grant numbers: AS-iMATE-107-11 and AS-105-TP-A03), Taiwan. H. T. Jeng also thanks the CQT-NTHU-MOE, NCHC and CINC-NTU, Taiwan for technical support.

[*]Corresponding authors.

wbsu@phys.sinica.edu.tw; jeng@phys.nthu.edu.tw

**Figure captions**

FIG. 1 (a) A differential *Z-V* spectrum with five FER peaks marked by numbers. Zero spectral intensity is indicated by a dashed line, revealing that the intensities of valleys (marked by arrows) around the peak of FER 2 are zero. Inset: a typical STM image of the MoS$_2$ surface. (b) Differential *Z-V* spectra with three, four, and six FERs. (c)–(f) Lorentzian fittings of FER 2 in (a) and (b). The linewidths extracted from the fittings change with the FER numbers in oscillation.

FIG. 2 (a) Energies of the high-order peaks in Figs. 1(a) and 1(b) versus $(n-1/4)^{2/3}$, showing the linear relationship, from which the electric fields for generating FERs of different numbers and vacuum level can be acquired. (b) Illustration of an STM tip consists of a base with a radius typically tens of nanometers and a protrusion (marked by an arrow). The electric field obtained from (a) appears at the dashed line normal to the surface for a range from o point to the classical turning point (cross) of the resonant electron. Because of this tip structure, the electric potential below the classical turning point in the lateral direction has a distribution curve. (c) Projected bulk band structure of MoS$_2$ obtained by DFT calculation for energy ranging from 1 to 12 eV above the Fermi level. Line 1 indicates the vacuum level. Lines 2 and 3 mark the band edges of a band gap. The energy diagram on the right-hand side shows that due to the band bending, the band edges and vacuum level in the bulk all move upward when approaching the surface. (d) RT time required by the resonant electron in the STM junction versus the electric field obtained from Fig. 2(a). (e) The resultant potential $U_r$ at the o point versus $\rho_0$ under electric fields obtained from (a).

FIG. 3 (a) One-dimensional square potential well with a width of 2$L$ for modeling the quantum trapping, where the energy *E* of the relaxed electron in the well is higher than the well depth $V_0$ and the wave function of the relaxed electron is the standing wave.

(b) Decay rate of the probability versus $\beta L$ that is proportional to the size of the well, revealing local minima for $E/V_0 = 0.5$ and $\gamma = 0.003$. Inset: maximum number of cycles $m_{max}$ for which the relaxed electron can remain in the well as a function of the decay rate for $N = 100$. (c) Average number of cycles versus $\beta L$ showing an oscillatory feature; the maxima occur exactly at the minima of the decay rate in (b). (d) Two electrons with opposite spins are emitted into the quantized state in the FER. (e) When one of resonant electrons in (d) is relaxed, its spin should be flipped to maximize the total spin according to Hund's rule.

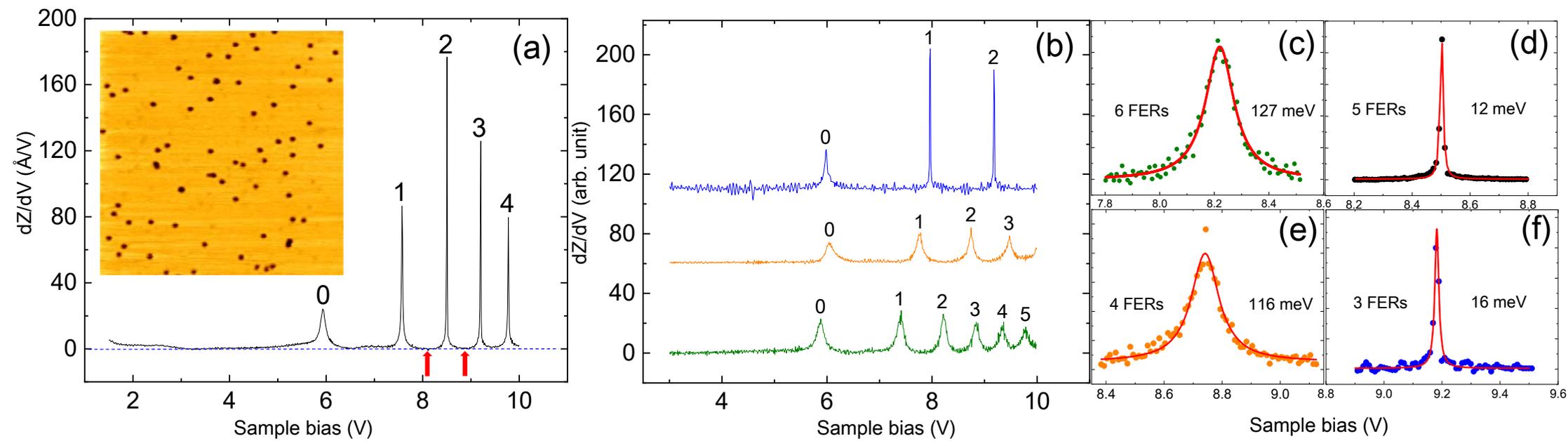

Figure 1

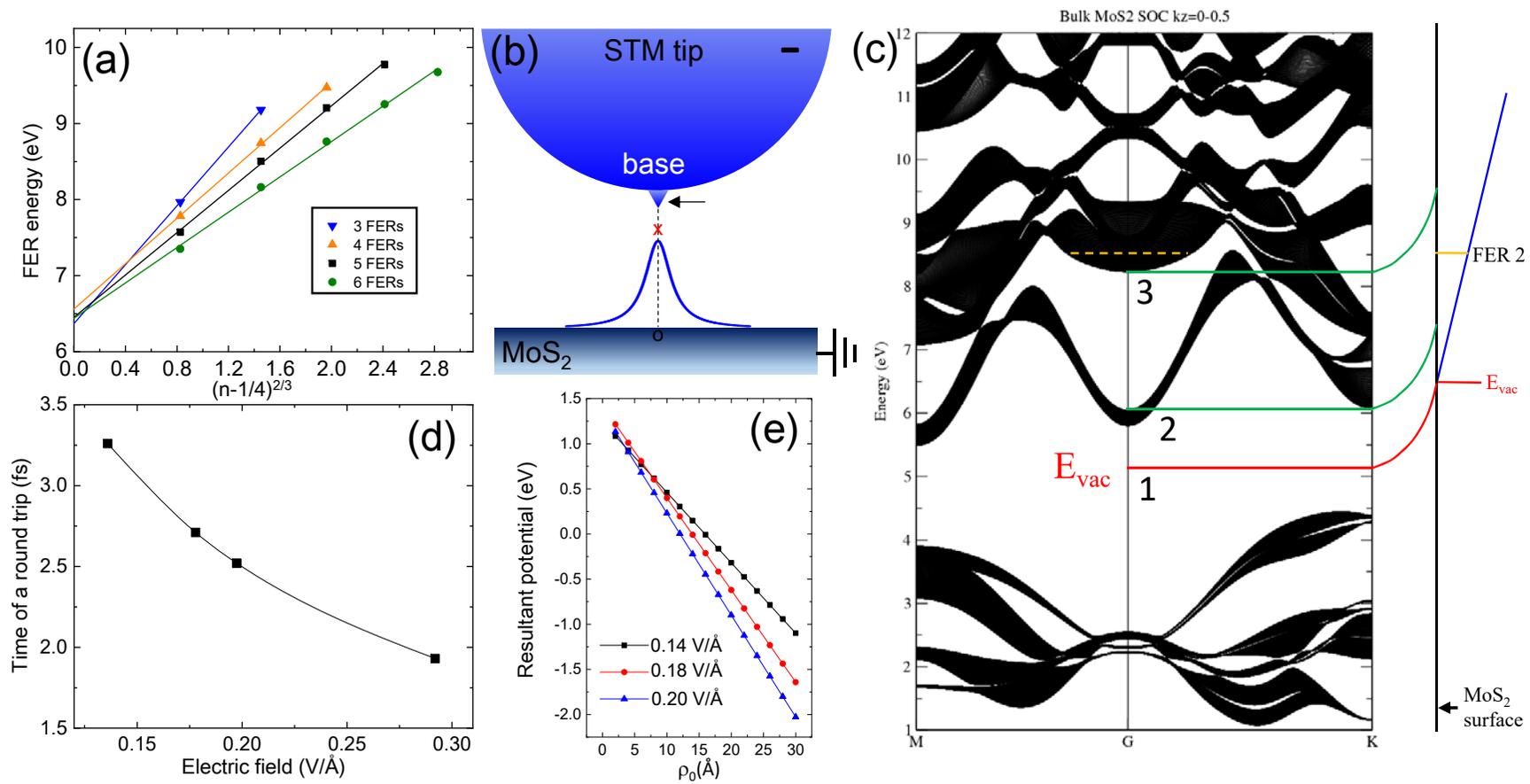

Figure 2

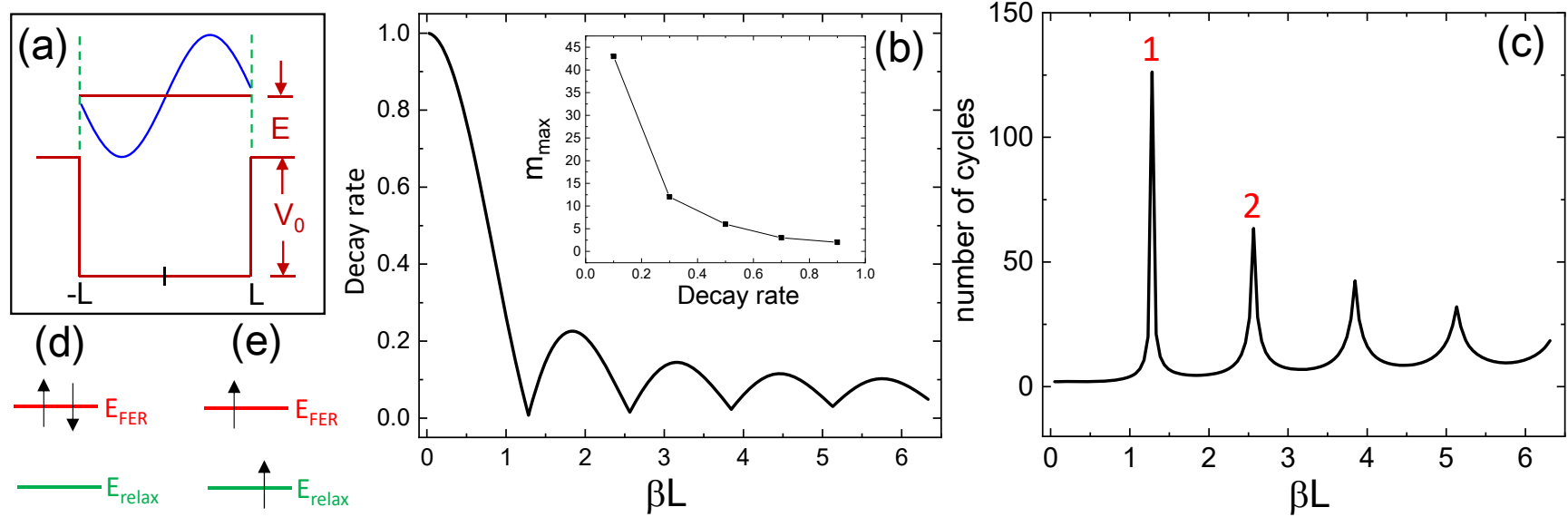

Figure 3

**Supplemental Material:**

# Quantum Trapping on MoS$_2$ through Lifetimes of Resonant Electrons: Revealing Pauli Exclusion Principle

Shin-Ming Lu,[1] Wei-Bin Su,[1*] Horng-Tay Jeng,[2,1,3*] Wen-Yuan Chan,[1] Ho-Hsiang Chang,[1] Woei Wu Pai,[4] Hsiang-Lin Liu,[5] and Chia-Seng Chang[1]

[1]*Institute of Physics, Academia Sinica, Nankang, Taipei 11529, Taiwan*

[2]*Department of Physics, National Tsing Hua University, Hsinchu 30013, Taiwan*

[3]*Physics Division, National Center for Theoretical Sciences, Hsinchu 30013, Taiwan*

[4]*Center for Condensed Matter Sciences, National Taiwan University, Taipei 10617, Taiwan*

[5]*Department of Physics, National Taiwan Normal University, Taipei 11677, Taiwan*

[*]Corresponding authors.

wbsu@phys.sinica.edu.tw; jeng@phys.nthu.edu.tw

## 1. STM measurements

A synthesis bulk Nb-doped MoS$_2$ sample (2D Semiconductor) was cleaved in air using an adhesive tape and then transferred into an ultrahigh vacuum STM operated at 78 K. FERs were observed by using the PtIr tip to perform *Z-V* spectroscopy under a current of 10 pA. Differentiation of the *Z-V* spectrum was executed using a numerical method to reveal FERs.

## 2. DFT calculations

The electronic structures were calculated using the projector augmented wave (PAW) approach as implemented in the Vienna Ab initio Simulation Package (VASP) based on the density functional theory (DFT). The Perdew−Burke−Ernzerhof (PBE) form of the generalized gradient approximation (GGA) was used for the exchange−correlation functional. The spin−orbit coupling was taken into account in the self-consistent field (SCF) calculations. For the bulk MoS$_2$ calculations, we used the 24 × 24 × 6 Monkhorst−Pack mesh for the k-point sampling within the 3D Brillouin zone. While 24 × 24 × 1 Monkhorst−Pack k-mesh over the 2D Brillouin zone was adopted for the 3-layer MoS$_2$ slab calculations with the vacuum thickness of 15 A well separating the slabs. The cutoff energy for the plane wave basis is set as 280 eV. The energy convergence threshold was set as $10^{-4}$ eV in the SCF calculations, while the energy convergence threshold was set as $10^{-3}$ eV for the structure optimizations.

## 3. Calculations for $\sigma(\rho)$ and $U_p(0)$

By using equation (3) in the text, the electric field $F_s(\rho)$ on the surface in the direction normal to the surface can be calculated to be $F/[1+(\rho/\rho_0)^2]$. Then, the density of the polarization charge $\sigma(\rho)$ is calculated as

$$\sigma(\rho) = \frac{\varepsilon_0(\varepsilon_r-1)F_s}{\varepsilon_r}, \qquad (1)$$

where $\varepsilon_r$ is the dielectric constant of MoS$_2$. Figure S1 shows the calculated density distribution of the polarization charge, displaying that the density is highest at the center and decreases with $\rho$.

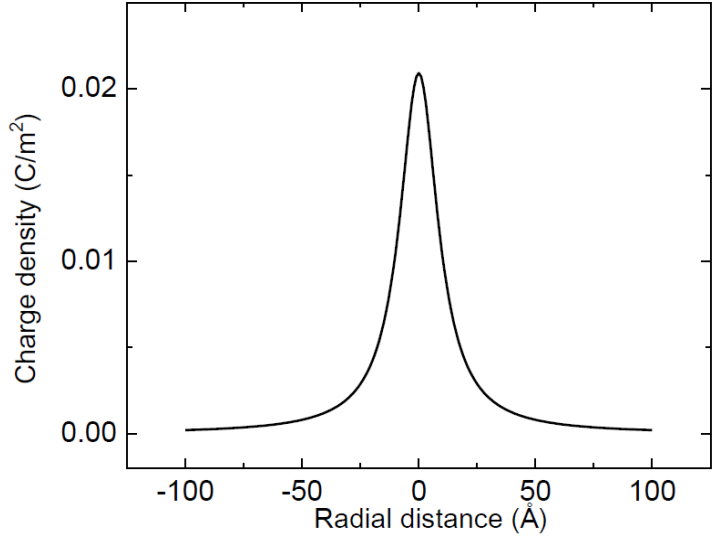

**Figure S1.** Calculated density distribution of polarization charge under the electric field of 0.178 V/Å and $\rho_0 = 10$ Å.

Because $\sigma(\rho)$ is acquired, $U_p(0)$ is calculated as

$$U_p(0) = -\int_0^\infty \frac{\sigma(\rho)2\pi\rho d\rho}{4\pi\varepsilon_0 \rho}. \qquad (2)$$

## 4. Polarization charge calculated by DFT calculation

Figure S2 displays the calculated valence band partial charge profiles along the z-direction of the 3-layer MoS$_2$ slab under uniform out-of-plane electric field ($E_z$) ranging from 0.0 to 0.3 V/Å. As can be seen clearly, the external electric field significantly modifies the surface MoS$_2$ layer valence charge profile by pulling electrons from the upper MoS$_2$ layer toward the lower MoS$_2$ layer, keeping the middle MoS$_2$ layer nearly neutral. As a result, the lower MoS$_2$ layer is negatively charged whereas the upper MoS$_2$ layer is positively charged, consistent with our expectation. The electric field induced electron migration grows monotonically with increasing electric field strength, forming the playground for manipulating the width and depth of the potential well on the MoS$_2$ surface by the electric field in the STM junction. We note here that the calculated charge density of the surface MoS$_2$ layer is of the same order of magnitude as the maximum charge density estimated in Figure S1.

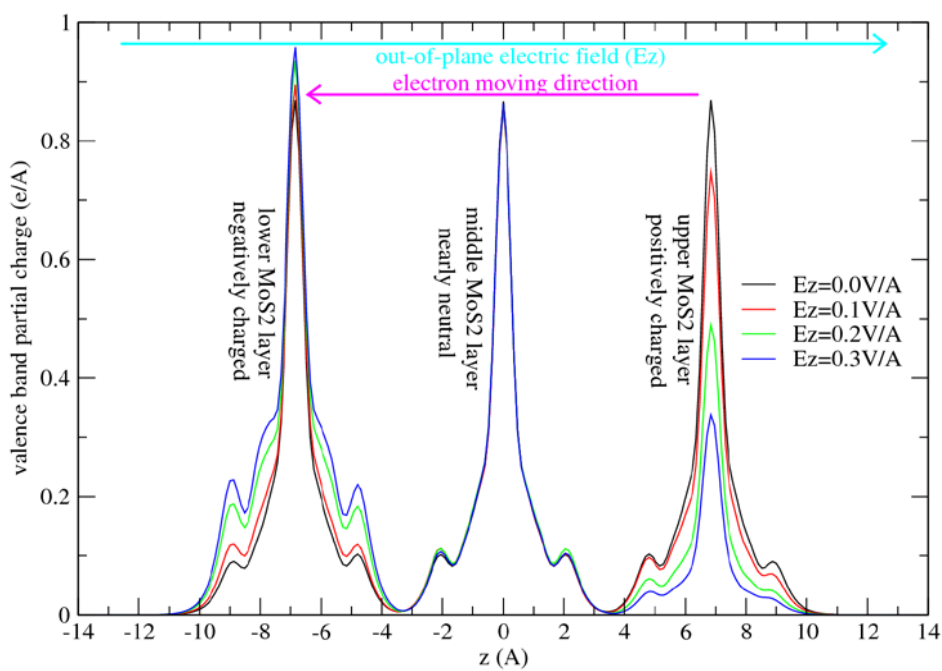

**Figure S2.** The charge profile of the valence band of MoS$_2$ slab under uniform out-of-plane electric field ranging from 0.0 to 0.3 V/Å. The external electric field significantly modifies the surface MoS$_2$ trilayer valence charge profile.